\documentclass[11pt]{article}

\usepackage[margin=1in]{geometry}
\usepackage[utf8]{inputenc}
\usepackage[british]{babel}
\usepackage{graphicx}
\usepackage{booktabs}
\usepackage{amsmath}
\usepackage[hidelinks]{hyperref}
\usepackage{needspace}

\usepackage[round,authoryear]{natbib}
\usepackage[none]{hyphenat} 
\usepackage{microtype}      
\title{Equity Bias: An Ethical Framework for AI Design}

\author{
	Mary Lockwood\\
	Independent Researcher\\
	United Kingdom\\
	\texttt{marylockwoodresearch@gmail.com}
}

\date{2026}

\begin{document}
	
	\maketitle

\begin{abstract}

Equity Bias is a philosophical and practical framework for building more equitable AI systems. It treats AI as an interpretive agent that constructs meaning and knowledge rather than as a neutral calculator. Grounded in Gadamerian hermeneutics and epistemic injustice theory, Equity Bias views bias not as an error to eliminate but as a reflection of whose knowledge is encoded into systems. Since all knowledge is situated, neutral systems are not possible. While traditional approaches aim to reduce or remove bias, Equity Bias instead makes bias transparent and contestable. 
Its three-stage AI Life Cycle (Equity Archaeology, Co-Creating Meaning, and Ongoing Accountability) turns these principles into practice. Equity Bias gives developers, researchers, policymakers, and communities affected by AI a method for examining whose knowledge a system encodes and who has the power to contest it. By placing the politics of knowledge at the centre of equitable AI design, Equity Bias offers a distinct approach to AI ethics.

\end{abstract}

\section{Introduction}

Artificial intelligence (AI) bias is one of the most complex ethical challenges of our time. Epistemology is the branch of philosophy concerned with how knowledge is produced and legitimised. As AI systems increasingly impact real-life decisions, the question of whose knowledge is encoded is critical. Addressing this challenge requires moving beyond purely technical fixes to confront the inherently political nature of knowledge production in AI systems. 

This paper introduces 'Equity Bias', a new design framework that views AI development through an epistemological lens. It treats bias not as an error to be eliminated, but as a meaningful feature of knowledge systems that must be openly negotiated. Equity Bias shifts development focus from unattainable neutrality to epistemic inclusion. Equity is achieved by ensuring that multiple knowledge systems inform and contest AI decision-making processes. By incorporating multiple worldviews at each stage of the AI life cycle, systems are better positioned to produce equitable outcomes for all of humanity. 

Current responses to AI bias often focus on statistical parity, fairness metrics, or data debiasing 
\citep{pessach2022, yang2024}. These approaches are valuable tools but typically treat bias as a mathematical problem based on the assumption that neutrality is possible and desirable. In contrast, Equity Bias understands bias as a reflection of underlying knowledge systems. 

Grounded in theories of epistemic injustice and Gadamerian hermeneutics, Equity Bias challenges the idea of value free AI. It positions AI systems as interpretive technologies that reflect the world and actively shape it. Equity Bias does not view all bias as negative or harmful for humanity. Society, for example, widely condemns human trafficking. Stripping AI systems of such normative biases in the pursuit of neutrality could, in some cases, produce ethically indefensible decisions and concrete harm. 

The term ‘epistemic gene pool’ refers to the range of data and perspectives that inform AI development. Deeper, more varied gene pools increase the adaptability and relevance of a system across contexts. Shallow pools heighten the risk of knowledge-limited AI systems that are blind to the realities they govern. To tackle harmful bias, Equity Bias argues for deliberately introducing more bias from multiple perspectives into systems. The goal is not to eliminate bias but to diversify and contest it. 

In practice, AI development often overlooks deeper epistemic questions. Issues impacting knowledge construction and inclusion remain unaddressed. Narrow data pipelines reinforce existing power imbalances between the people who control technologies and those impacted. This limits the ability of AI to produce equitable outcomes. 

As a framework, Equity Bias has clear implications for practice. For developers, it demands new approaches to sourcing and annotating data. For researchers, it invites interdisciplinary engagement with knowledge production. For policymakers, it expands the regulatory lens from 
statistical outputs to the epistemic processes behind them. By broadening the epistemic gene pool, Equity Bias supports the development of smarter, more accountable AI systems.

\section{Theoretical Foundations}

Understanding why current approaches to AI bias fall short requires looking beyond technical fixes. It means examining the philosophical roots of bias itself. Too often, AI bias is treated as a problem that can be solved with cleaner data or stronger algorithms \citep{gichoya2023,ferrara2024}. This approach ignores the social and historical forces that shape what counts as knowledge \citep{zajko2022}. AI bias reflects human bias. It is complex, messy, and deeply embedded in social structures. Recognising this truth moves the challenge from eliminating bias to critically engaging with the knowledge systems AI encodes.

\subsection{Epistemic Injustice}

Epistemic injustice refers to harm done to individuals in their capacity as knowers. \citet{fricker2007} identifies two primary forms, testimonial injustice and hermeneutical injustice. Testimonial injustice occurs when prejudice leads a listener to unfairly discount the credibility of a speaker. For example, a low-income manual labourer may be viewed as a less reliable witness in court than a high-income suited professor.

Hermeneutical injustice arises when gaps in shared understandings prevent people from interpreting and expressing certain experiences. For example, women in the 1960s faced sexual harassment at work but lacked the language to describe it, leading to dismissal or misunderstanding \citep{fricker2007}. Both testimonial and hermeneutical injustice serve as a warning of how AI systems can replicate and scale forms of epistemic harm. 

AI systems commit testimonial and hermeneutical injustice in digital contexts. Testimonial injustice emerges when models trained on dominant group data systematically diminish marginalised voices. For example, recruitment algorithms have replicated gender and racial biases by ranking minority candidates lower than non-minority candidates despite equal qualifications \citep{chen2023}. Existing social disparities are reinforced by how AI encodes assumptions about whose voices matter. 

Hermeneutical injustice in AI arises when systems lack the conceptual resources to interpret marginalised experiences. Models trained on culturally dominant data inherit blind spots that obscure non-normative perspectives. In fields such as medicine, this can cause serious harm. 

Diagnostic tools, for example, have misdiagnosed women with liver disease at twice the rate of men due to training data being skewed toward male cases \citep{straw2022}. Limited diversity in training data accounts for why AI often fails to detect dermatological conditions on darker skin compared to lighter skin \citep{lopezperez2025}. By excluding underrepresented experiences from what is considered knowable, AI systems can produce false negatives and positives that adversely impact health outcomes. Expanding the field of knowledge input sources improves the accuracy of knowledge outputs. 

Generative AI can perpetuate hermeneutical injustice by reflecting and reshaping dominant interpretive standards. Systems that are trained on vast internet datasets inherit embedded cultural biases \citep{kay2024} such as attitudes towards mental health. As generative models increasingly function like search engines \citep{lindemann2024}, they shape what knowledge is visible and credible. AI systems can marginalise knowledge at scale, shaping whose voices 
people view as legitimate and illegitimate. 

Epistemic injustice in AI reflects deep, systemic social inequalities. Tackling this issue requires more than just better data or improved models. It demands rethinking how knowledge is defined, whose voices are deemed important, and what it means to build systems that value multiple ways of knowing. 

\subsection{Hermeneutics}

Hermeneutics is the philosophy of interpretation. It offers crucial tools for understanding AI as a meaning-making technology rather than a neutral calculator. Originally developed to interpret texts, hermeneutics can help us to understand AI systems as interpretive agents. Gadamer’s work is especially relevant, as AI increasingly processes human language and behaviour to generate meaning and shape decisions. 

\citet{gadamer2004} challenges the idea of neutral interpretation. He argues that all understanding is shaped by prejudices, understood as the pre-judgements, assumptions, and prior experiences that inform an initial grasp of a situation or text. Interpretation is always historically and culturally situated. Gadamer’s concept of the ‘fusion of horizons’ describes how new knowledge and understanding emerge through dialogue between different perspectives. When existing sources of knowledge collide, new knowledge is formed from the fusion. Every perspective is shaped by unique experiences and backgrounds.  In the context of AI, this fusion suggests systems could benefit from incorporating multiple knowledge sources rather than a single dominant perspective.

The interpretive capabilities of AI require active human judgement to ensure that they support rather than distort ethical decision-making \citep{bakdash2023}. Interpretation by AI is inherently contextual and dynamic as opposed to value free. For a successful evaluation of AI, we must consider not only what it computes but also how it interprets the world and how humans interact with its outputs.  

Drawing from this lens, Equity Bias departs from popular consensus within responsible AI movements and does not view bias as automatically negative. Pre-judgements can be positive and negative. They are the necessary conditions for understanding that stem from our historical and cultural experiences. AI systems cannot escape interpretive bias and do not present purely objective analyses. This reframing invites deeper reflection on how AI systems shape social and political meaning. 

Bias and prejudice are not a defect but a necessary foundation for interpretation \citep{gadamer2004}. Erasing AI bias is not the ethical challenge because it is impossible. The challenge is to reflect on system bias transparently and engage with it critically. Ignoring this leads to the false hope of neutrality and perpetuates exclusion. Embracing bias opens the door to dialogue and pluralism in AI. 

The field of medical humanities provides concrete examples of hermeneutic principles in practice. In clinical settings, power imbalances often privilege the doctor’s voice over the patient’s voice. Yet patient stories reveal forms of understanding that clinical data alone cannot capture \citep{frank2013}. Traditional medical hierarchies often exclude and devalue the patient’s voice. This diminishes the quality of understanding and decision-making, potentially harming patient-centred care \citep{cote2024}. Mainstream ways of knowing often overlook or dismiss alternative perspectives, limiting what is considered as valid knowledge. 

When AI models are trained on dominant cultural narratives, they risk reinforcing structural inequalities and excluding other ways of understanding the world. In AI contexts, algorithmic systems risk embedding and increasing exclusions \citep{noble2018}. Without intentional inclusion, AI may reinforce patterns of exclusion that have lasting societal consequences. 

AI’s reliance on dominant cultural data and unexamined assumptions often narrows its capacity to understand. To foster fairer AI, we need hermeneutic accountability, which means treating AI as a participant in meaning-making rather than a neutral tool. This involves building systems that recognise their interpretive positions, invite critique, and actively engage with the communities they affect. Bias cannot be erased, but it can be critically negotiated through transparency and inclusion.  

\subsection{Limits of current frameworks to address AI bias}

Dominant AI fairness frameworks often treat bias as a measurable error that can be mitigated through better data or debiasing algorithms \citep{jui2024}. Common metrics such as demographic parity (equalising outcomes across groups) and equalised odds (equalising error rates) focus primarily on statistical fairness \citep{buijsman2023}. These approaches are valuable for identifying disparities, yet they may overlook whose perspectives shape these definitions and which outcomes are prioritised.

Models trained predominantly on data from dominant groups may struggle to capture underrepresented experiences. This limitation is not merely technical. It reflects a deeper epistemic and hermeneutical challenge. Fairness metrics alone cannot fully address structural forms of bias, particularly when bias is framed solely as a correctable error rather than as an inherent feature of how knowledge is represented.

\citet{wang2023} distinguish between aleatoric discrimination arising from data uncertainty, and epistemic discrimination rooted in model design and knowledge representation. This distinction highlights how bias can manifest at multiple levels within AI systems. While current frameworks can help identify and mitigate some forms of bias, they may inadvertently reinforce historical exclusions if applied without attention to power, perspective, and context. Such observations suggest the limits of purely objective algorithmic approaches to complex social issues.

Moving forward, AI research should focus on representing knowledge inclusively, ensuring that multiple perspectives are incorporated to reduce systemic biases and better align technical solutions with social realities.

\subsection{Epistemic Pluralism}

Epistemic pluralism acknowledges that knowledge is always situated and enriched by diverse perspectives \citep{fricker2007}. It challenges AI to move beyond mere calculation toward interpretation and sense-making. Instead of chasing neutral algorithms, this approach embraces multiple ways of knowing and rejects the myth of objectivity that often masks dominant views. AI systems built on these principles function as interpretive agents, open to challenge and revision.  

Transforming epistemic pluralism into practice is challenging but achievable. Participatory design offers concrete ways to implement epistemic pluralism in AI systems. Involving affected communities in the design process can help tackle the issue of bias without relying solely on sensitive data \citep{veale2017}. Educational AI systems that incorporate diverse epistemic perspectives are more likely to support inclusion and decrease marginalisation \citep{song2024}. Inclusion matters because it expands the range of knowledge and experiences that inform system design. Dialogue and reflexivity should become core features of AI systems. 

Embedding pluralism requires systemic changes that extend beyond technical design. It demands institutional and cultural changes that value different ways of knowing to increase the trustworthiness of AI decision making. AI must be transparent about its interpretive limits and responsive to communities so that it reflects a plurality of needs. However, pluralism also presents its own challenges. Navigating conflicting perspectives and complex governance requires reflexive and participatory decision-making supported by ongoing dialogue and adaptable frameworks. 

\subsection{Statistical Fairness Limitations}

Current AI fairness research emphasises metrics like demographic parity and equalised odds that promise precision but oversimplify social complexity. By focusing narrowly on algorithmic fixes while excluding community voices, these approaches risk becoming superficial compliance exercises that entrench the very inequalities they claim to address. Bias in AI is not just a numerical problem. It reflects systemic power relations embedded in data, design, and deployment. 

Identity is fluid and intersectional. People experience overlapping forms of disadvantage that resist simple categorisation \citep{crenshaw1989}. Hiring algorithms optimised for gender parity may still discriminate against women of colour because treatment of race and gender as isolated variables misses their combined impact \citep{drage2022, kelan2024}. The result is systems that meet diversity targets while reproducing exclusion through invisible intersections. 

Trade-offs between fairness metrics such as demographic parity and equalised odds force designers to prioritise one conception of fairness over another \citep{chouldechova2017}. Yet these decisions are typically made by technical teams without democratic input or transparency about their broader social implications. Statistical approaches detach lived experience from numeral proxies. This enables what \citet{birhane2021} calls ‘fairness theatre,’ which describes systems that meet diversity metrics but fail to understand or meaningfully serve marginalised communities.  

The myth of clean data assumes neutrality can be engineered through preprocessing and balancing techniques. However, debiasing often strips out information needed to understand discrimination patterns \citep{simson2024}. Removing race or gender from hiring datasets can obscure systemic inequities while allowing bias to persist through proxy variables like postal codes or educational institutions. These ‘fair’ systems reproduce exclusion through hidden pathways while appearing impartial. 

Data-centric approaches embed power relations in technical decisions about representation and categorisation. Systems typically rely on pre-existing institutional categories and sampling frameworks, with limited community input into how identity should be defined or measured \citep{benjamin2020}. Current AI development processes thus create systems that serve organisational efficiency over community recognition. Technical balance cannot address exclusion rooted in whose knowledge counts as legitimate, which is why data driven fairness cannot be relied on to remove systemic inequities. 

\subsection{The Neutrality Myth and Systematic Harm}

The neutrality myth refers to the false belief that technology can be ideology-free \citep{noble2018}, yet data is never neutral \citep{catania2023}. By framing political choices as technical inevitabilities, this myth enables systematic harm through ethics-washing and epistemic violence. 

Ethics-washing occurs when organisations adopt fairness metrics as performance rather than reform. Companies may demonstrate algorithmic auditing compliance while avoiding questions about surveillance, labour exploitation, or democratic accountability. Fairness becomes a checkbox that legitimises harmful systems by creating an appearance of ethical consideration without substantive transformations. 

The UK's OFQUAL algorithm demonstrates this dynamic. Used to predict A-level results during the Covid-19 pandemic, it systematically downgraded students from lower-income schools while being defended as fair and objective \citep{madaio2021}. Only widespread protest forced acknowledgment of its discriminatory impact, revealing how ‘neutral’ systems can entrench structural inequality under claims of objectivity. 

Technical solutionism treats centuries of structural inequality as engineering puzzles with optimal solutions. It depoliticises AI development by suggesting expert optimisation can resolve complex social problems without democratic deliberation \citep{jungherr2023}. Claims of universal design can unintentionally centre dominant perspectives and marginalise alternative viewpoints. Together, these dynamics create systems that embed values while claiming universal validity. 

\subsection{Critical Alternatives and Implementation Gaps}

Emerging critical approaches challenge mainstream assumptions but remain largely peripheral to practical AI development. Data Feminism explores how power shapes data systems and calls for intersectional analysis \citep{dignazio2020}, effectively surfacing politics embedded in technical decisions. However, its primary focus on data limits engagement with broader questions of system architecture and deployment contexts. 

Design Justice promotes democratic participation and accountability in technology development \citep{costanzachock2020}, providing compelling alternatives to top-down design. While it has influenced some mainstream practices, integrating it within the bureaucratic and commercial constraints of large-scale AI projects remains challenging. 

Despite these valuable insights, adoption remains limited. They confront core assumptions about objectivity, expertise, and technical sufficiency that underpin current AI development. Their practical influence remains constrained by the absence of concrete methodologies that can operate within existing institutional frameworks while maintaining critical commitments.

\subsection{The Need for Equity Bias}

Current approaches fall short because they address symptoms rather than underlying causes. Statistical fairness targets outcomes without examining the processes that produce bias. Data-centric models offer technical fixes for structural issues, while critical alternatives provide important perspectives but often lack clear methods for real-world application. 

Equity Bias contributes a distinct framework grounded in hermeneutic philosophy. Existing participatory frameworks like Design Justice and Data Feminism identify who is excluded from data systems. Equity Bias provides a lens for understanding how AI systems function as interpretive agents that actively construct meaning. This hermeneutic turn reframes bias from error to epistemic position. Bias is not automatically harmful. The challenge is making interpretive stances transparent and contestable. Regulatory focus shifts from pursuing impossible neutrality to negotiating whose interpretations shape AI’s understanding of the world. 

Equity Bias offers a new direction. It views bias not as a flaw to eliminate, but as a reflection of embedded values that must be made transparent and accountable. By combining normative commitments with practical design methods, it supports meaningful change within existing institutions.

\section{Equity Bias: A Practical Framework}

Equity Bias is a framework for designing and evaluating AI systems that challenges who gets to define knowledge and meaning. Rather than treating bias as a technical flaw to be corrected, it frames bias as a lens revealing whose knowledge is included and excluded. This perspective shifts the focus from technical correction to ethical interpretation. 

Bias, in this framework, can be both negative and positive. It is a window into power revealing whose knowledge shapes systems. Efforts to ‘debias’ systems often mask dominant assumptions rather than confront them. Equity Bias pushes beyond the myth of neutrality and creates space for perspectives that may challenge institutional comfort or mainstream norms. 

This shift reshapes how AI systems interpret the world. Consider the experiences of former child soldiers. Their stories may clash with social norms or institutional narratives, but they are essential to designing systems that respond meaningfully to real conditions. An anti-radicalisation or Post Traumatic Stress Disorder (PTSD) support tool must account for such lived realities to be as effective as possible. Systems that overlook them risk misrepresentation or harm. 

Equity Bias is not just about fairness or representation. It rethinks who holds the power to define what matters. By embracing disagreement and complexity rather than smoothing them away, it encourages more accurate, accountable, and socially grounded AI.

\subsection{Guiding Principles}

The Equity Bias framework rests on four guiding principles that reorient how AI systems handle knowledge and inclusion. It begins with the core assertion that AI is never neutral. Every dataset, model, and interface reflects visible and hidden values, assumptions, and histories. To make AI more accountable and prevent existing inequalities from being reinforced, we must surface and rethink these foundations. 
 
\vspace{0.5em}
\textit{Knowledge Is Situated} 
\vspace{0.5em}

All knowledge comes from somewhere, including people, places, and power structures. Every AI system reflects assumptions about what is true, useful, or valuable. Such assumptions often reflect dominant worldviews while sidelining others. This can lead to exclusion and misrepresentation, causing models to malfunction for certain groups. Relying on incomplete or biased data can produce costly financial errors and harms. 

A language model trained on Western academic texts may enforce a specific worldview as if it were universal. Likewise, Chinese AI models shaped by state censorship embed different, but equally powerful, boundaries on what can be said or known. These systems do not merely process data. They define and create what counts as knowledge. AI systems can reinforce dominant power structures by shaping whose knowledge is seen as legitimate. 

\needspace{5\baselineskip}
\vspace{0.5em}
\textit{Knowledge Is Contested}
\vspace{0.5em}

Equity Bias does not treat all knowledge as interchangeable but challenges who decides what counts as valid or real. When scientific, experiential, and cultural knowledge systems intersect, clashes of perspective can expose hidden assumptions within each. 

Health AI that integrates both clinical research and community understandings of mistrust does not dilute scientific rigour. It becomes better equipped to engage with and address mistrust. Agricultural AI designed solely for industrial farming may overlook Indigenous practices rooted in alternative understandings of land and ecology. Combining both knowledge systems may lead to more effective and sustainable farming outcomes. The tension between different ways of knowing is a source of insight that has the potential to solve complex problems.  

Equity Bias acknowledges that reality is not singular or fixed. What is accepted as fact today may be considered fiction in the future. Even beliefs labelled ‘incorrect’ by institutions often reveal deeper truths about risk, trust, and exclusion. The goal is not to resolve differences by choosing a correct worldview but to learn from diverse perspectives and expand who shapes AI’s understanding as opposed to narrowing it. 

\vspace{0.5em}
\textit{Inclusion Requires Shared Knowledge-Making} 
\vspace{0.5em}

‘Inclusion’ refers to those directly affected by AI systems. It is not a box to tick at the end of design but a process that starts with who helps define the problem. Equity Bias shifts inclusion from token representation to meaningful participation. AI development becomes a process of shared learning and negotiation where different forms of knowledge (e.g. scientific, cultural, experiential) are all considered. 

This does not mean redistributing full decision-making power. For example, patients should not decide what surgical procedures other patients receive. However, their lived experiences of health and services can reveal critical care blind spots that may reduce harm to others down the line. When AI recognition systems misinterpret cultural dress, the problem is not only technical but also a failure to understand different ways of seeing. Inclusion challenges embedded power dynamics by ensuring marginalised voices help shape how AI systems interpret the world. 

Participatory inclusion recognises technical expertise and lived experience as complementary sources of knowledge. In disaster response, emergency responders may understand protocols, whereas locals may know which streets have flooded or where vulnerable individuals are stranded. Systems designed with input from both perspectives can be more effective, accountable, and trusted by the communities they serve. 

Shared knowledge-making is at the heart of Equity Bias’ commitment to ethical AI. It grounds technology in the lived realities of those it impacts, rather than relying on abstract ideals alone. 

\vspace{0.5em}
\textit{Systems Must Stay Open to Change} 
\vspace{0.5em}

Knowledge should not be treated as fixed or unchanging. As communities evolve and new insights emerge, AI must remain flexible technically and ethically. The ability to question foundational assumptions about whose knowledge matters is valid and valuable. Without this adaptability, systems risk becoming disconnected from the people they serve and entrenching existing inequalities. 

Even well-intentioned tools can unintentionally reinforce outdated beliefs or practices if they fail to incorporate multiple evolving viewpoints. The need for adaptability is particularly clear in AI media applications where new voices, genres, and cultural shifts constantly reshape what knowledge is valued and how it is categorised. By staying open to challenge and revision, AI systems are more likely to produce accurate outputs. 

\subsection{Rethinking Key Assumptions}

To put these ideas into practice, Equity Bias proposes six foundational shifts (see Table 1) in how AI systems understand and structure knowledge. Assumptions underpinning these systems are as significant as their outputs. This means expanding the design lens beyond solely technical fixes. Technical choices and ethical considerations are deeply intertwined. Addressing bias requires both rigorous methods and critical reflection on the values, assumptions, and power structures embedded in AI systems. 

\begin{table}
  \caption{Foundational shifts in AI design}
  \label{tab:foundational-shifts}
  \begin{tabular}{p{0.45\linewidth} p{0.45\linewidth}}
    \toprule
    \textbf{From} & \textbf{To} \\
    \midrule
    Fixing bias as an error & Valuing multiple ways of knowing \\
    Claims of objectivity & Transparent reflection on who shapes the system \\
    Standard fairness metrics & Recognition of historical and social context \\
    Accuracy as a universal goal & Contextual accuracy informed by lived experience \\
    Assumptions of `clean' data & Tracing the origins and context of data \\
    Assumptions of generalisability & Designing context-aware, locally grounded systems \\
    \bottomrule
  \end{tabular}
\end{table}

These changes prompt deeper questions. Who decided what matters in this system? Who is missing? What alternatives were overlooked?  

Accuracy is often defined by how closely a model matches labels in a benchmark dataset. However, what happens if that dataset erases human diversity? Voice assistants may meet technical standards and still fail with certain accents because they were trained on a narrow idea of whose speech is normal. It is not just a glitch. It reflects which voices were included and which were excluded. 

Likewise, the idea of clean data is misleading. All data is shaped by the context in which it was collected. For example, financial data may carry the legacy of redlining or loan discrimination. If the stories of data are ignored, systems risk quietly reproducing societal harms. Tracing where data comes from is not optional. It is foundational to ethical design. 

Another common assumption is generalisability, the belief that a model that performs well in one setting will work everywhere. Knowledge is shaped by context. AI tools built on India’s legal or educational norms may fail in different countries. Equity Bias treats differences not as obstacles but as core design requirements. Context-aware systems must be built with local knowledge from the start rather than retrofitted later. 

\subsection{Applying the Framework}

\begin{figure}[ht]
    \centering
    \includegraphics[width=0.5\linewidth]{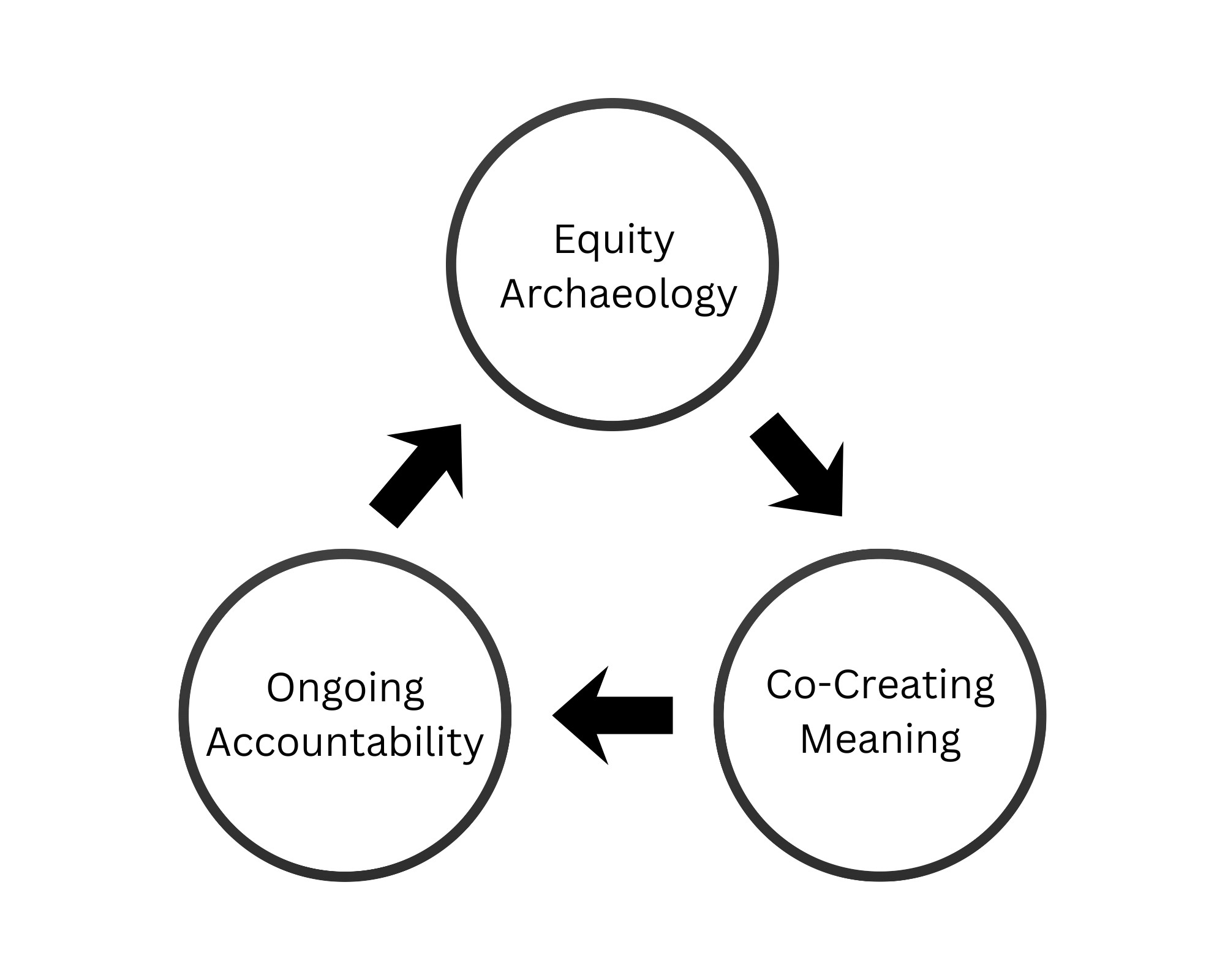}
    \caption{Equity Bias AI Life Cycle}
    \label{fig:equity-bias-ai-life-cycle}
\end{figure}

Equity Bias is most effective when embedded throughout the entire AI life cycle. Rather than a checklist applied at the margins, it fosters continuous ethical reflection and inclusive knowledge practices at every development stage. This section introduces the Equity Bias AI Life Cycle (see Figure~\ref{fig:equity-bias-ai-life-cycle}) outlining a conceptual approach to iterative, accountable, and context-sensitive system building. Together, these phases form a feedback loop that continually re-examines and aligns AI systems with multiple perspectives.

\needspace{5\baselineskip}
\vspace{0.5em}
\textit{Phase 1: Diagnosis (Equity Archaeology)}
\vspace{0.5em}

The first phase centres on uncovering the worldviews, exclusions, and value judgements built into system design. This stage treats AI as a social construct shaped by choices about whose knowledge matters and who decides what success looks like rather than simply a technical tool. Developers must actively map the epistemic gene pool and identify all stakeholders affected by the system. Key decisions regarding data selection, goal setting, and underlying assumptions must be examined.  

For example, many pain management tools are based on data from male patients, as women have been historically excluded from medical research. This means AI may misjudge women’s pain levels or recommend fewer effective treatments. Uncovering whose experiences have been excluded from training data can help create more equitable and effective tools.  

Equity Archaeology is not about assigning blame. Assumptions can be enabling and limiting. It is about making the normative foundations of a system visible. Insights gained in this phase do more than guide technical fixes. They lay the groundwork for more grounded and equitable collaboration in subsequent design phases. 

\vspace{0.5em}
\textit{Phase 2: Design (Co-Creating Meaning)} 
\vspace{0.5em}

Building on insights from diagnosis, the design phase shifts focus to collaboration. Instead of treating communities as sources of feedback or data, co-creating meaning invites them to shape how the system works from the outset. This includes co-framing the problem, identifying priorities, and defining what success should look like. 

Design is not simply about implementing technical solutions. It is a space of negotiation. Stakeholders often hold conflicting values and goals. Tensions should be visible rather than erased. By embracing multiple perspectives and resisting premature consensus, the design phase preserves complexity. This helps to ensure that the final system better reflects varied needs and avoids oversimplified, one-size-fits-all outcomes. 

For example, an AI tool designed to support domestic abuse prevention may assume users have safe access to police or shelter services. Yet these may not be viable options for many people. Without input from affected communities, the system could recommend unsafe actions or fail to identify key signs of risk. Involving survivors and local advocates helps build tools that reflect multiple needs and avoid harmful assumptions.

By grounding design in shared meaning-making, systems are more aligned with real-world needs. It also lays the foundation for evaluation practices that reflect community defined goals, along with institutional defaults. Done well, co-design builds trust, deepens accountability, and leads to more adaptive and responsive systems. 

\vspace{0.5em}
\textit{Phase 3: Evaluation (Ongoing Accountability)}
\vspace{0.5em}

The evaluation phase redefines success by maintaining system responsiveness over time. Evaluation is not conceived as a one-time checkpoint but as a living, ongoing process that evaluates whether AI systems serve the needs of those they affect. This requires mechanisms for regular community input, assessment of how the system adapts to shifting contexts, and redistribution of decision-making power, as necessary. 

For example, an adverse weather response system might initially perform well but lose relevance as climate patterns and community needs evolve. A standing community board empowered to trigger redesigns can help maintain alignment between the system and its users, ensuring long-term accountability and adaptability. 

The diagnosis, design, and evaluation phases form a continuous, iterative loop. Insights gained during evaluation feed back into new rounds of diagnosis and co-design, enabling AI systems to evolve in tandem with the communities they serve. For example, an education platform may begin with standard assessments. Through ongoing community feedback, it might identify cultural biases and co-create alternative evaluation methods that further adapt as student needs change. This cycle fosters systems that learn with, not just about, the people impacted. 

\subsection{Implementation Challenges}

Equity Bias asks organisations to rethink how they develop AI. It challenges common assumptions about efficiency, consensus, and success. Core shifts are needed to integrate multiple perspectives meaningfully. 

AI success is often measured by how quickly and widely a system can be deployed. Equity Bias reframes success in terms of adaptability, meaningful integration of multiple perspectives, and responsiveness to the evolving needs of those affected. Achieving these outcomes requires different skills, evaluation approaches, and timelines compared to conventional development models. 

For example, an urban mobility AI might be celebrated for quickly scaling across a city’s transport network. Yet if it was trained primarily on the travel patterns of able-bodied commuters, it may unintentionally restrict accessible routes for people with disabilities. By expanding how success is defined, Equity Bias helps build systems that are not only efficient but also accurate, trusted, and relevant in the long term.

\subsection{Managing Multiple Perspectives}

Multiple perspectives are not limited to cultural or ethnic difference. Equity Bias values insights from those whose experiences sit both within and outside dominant narratives. This includes economically disadvantaged groups, rural communities left out of the tech loop, and others whose knowledge is often overlooked. The goal is to surface neglected perspectives and design systems that are richer, more grounded, and more responsive. 

Diversity of thinking brings complexity, which is a strength. Disagreement signals that different ways of understanding the world are being taken seriously. The aim is not to force consensus, but to hold differences in active conversation. This helps AI systems reflect a broader range of needs and realities. 

Importantly, not all disagreements stem from cultural difference. People from the same community often hold conflicting views. What matters is building systems that can recognise, hold, and work with those tensions. While this may be unfamiliar to organisations used to streamlined processes, it makes systems more flexible, accurate, and better suited to real-world complexity. 

\subsection{Organisational Shifts and Institutional Resistance}

Implementing Equity Bias is not a quick fix. It requires long-term change. Leadership must be willing to value multiple inputs when defining problems and interpreting outcomes, even if it challenges familiar ways of working. Organisations that embrace ongoing learning and adaptation are better equipped to benefit. This means building formal consultation channels, setting up regular feedback loops, and hiring or training staff who can work across different ways of knowing. These are not just add-ons. They represent structural shifts in how organisations understand and manage knowledge. Over time, such changes can improve efficiency by helping prevent costly mistakes and misaligned decisions. 

Approaches that may alter structures are often resisted by organisations. Equity Bias may unsettle established power dynamics by challenging conventional definitions of expertise and broadening participation. Successfully addressing this resistance involves cultivating new communication channels, developing shared vocabularies, and building the capacity to collaborate across disciplines and organisational boundaries. Meaningful implementation requires recognition and management of cultural shifts. 

Under Equity Bias, scalability means adaptability rather than uniformity. Instead of applying the same model everywhere, organisations tailor shared principles to local needs. This happens through practices like stakeholder mapping, context-specific consultation, and flexible evaluation methods that can evolve over time. 

These approaches are not imposed top-down. They grow iteratively, shaped by ongoing engagement and locally grounded knowledge. Such flexibility takes time and effort, especially for smaller organisations who may not be able to afford the resources required to implement the approach. Development of scalable tools and support structures could make context-sensitive design more feasible at different scales.

\subsection{Public Health Case Study}

AI-driven disease surveillance systems typically rely on official data such as hospital admissions and lab results. Informal sources of data such as social media are often excluded. Institutional data is treated as more reliable, while community-generated data is framed as noisy or anecdotal. For example, dengue forecasting models in Southeast Asia commonly use quantitative data \citep{colongonzalez2023}. However, social media indicators frequently provide earlier outbreak signals than formal systems \citep{ramadona2019}. Narrow data focus limits early detection and reduces the adaptability of public health responses. To address this, systems need to incorporate diverse forms of intelligence. In this context Equity Bias offers a solution by broadening the epistemic base of health surveillance AI. 

The exclusion of informal data sources does not merely delay outbreak detection. It reveals a deeper failure mode in current surveillance systems. When uncertainty is high and official reporting lags, models trained to favour institutional data risk degraded operational performance. This is not a technical oversight but a structural blind spot produced by assumptions about whose knowledge is credible. As a result, systems perform worst precisely when early, context-rich signals are most needed.

\begin{itemize}
    \item \textbf{Equity Archaeology}: Examine whose knowledge shapes detection systems. Are outbreak alerts based only on institutional data, or do they incorporate insights from frontline workers, local media, and affected communities?
    \item \textbf{Co-Creating Meaning}: Build models in partnership with public health workers, community organisations, and media monitors. This collaboration helps distinguish between misinformation and meaningful early signals which can improve contextual understanding and response accuracy.
    \item \textbf{Ongoing Accountability}: Establish feedback loops that compare AI forecasts to how outbreaks unfold on the ground. This enables models to adjust to shifting local conditions and evolving public perceptions of risk.
\end{itemize}

In public health contexts, the consequences of epistemic exclusion are not abstract. Delayed detection, misallocated resources, and ineffective communication can directly shape morbidity, mortality, and public trust. Surveillance systems that fail to account for how knowledge is produced and interpreted risk reproducing the vulnerabilities they aim to mitigate. By treating epistemic diversity as a design requirement rather than a supplementary feature, Equity Bias supports forms of AI governance better suited to uncertainty, crisis response, and accountability.

\subsection{Defence Case Study}

Militaries worldwide face critical challenges in digital transformation, including how AI shapes decision-making under complex, rapidly evolving conditions. Epistemic narrowing is the progressive narrowing of knowledge sources deemed legitimate. It is effectively the opposite of Equity Bias, which seeks to expand the range of perspectives informing AI systems. Epistemic narrowing not only limits insight but raises the risk of failure in fast-moving situations, where over-reliance on narrow data sources can result in serious misjudgements and harm. 

Defence AI often mirrors institutional hierarchies, favouring senior analysts, standard doctrines, and establishment intelligence sources. This can undervalue non-traditional intelligence contributors whose insights could strengthen situational awareness and mission success. When these perspectives are excluded from system design and data inputs, AI tools risk reinforcing existing blind spots rather than improving decision-making. 

Implementing Equity Bias in defence contexts requires navigating legitimate operational constraints. Classification requirements, time sensitive decision-making, and personnel security levels present real boundaries around information sharing. Equity Bias does not advocate for compromising operational security or eliminating command hierarchies. 

Instead, Equity Bias challenges organisations to examine whether epistemic exclusions stem from genuine security needs or from institutional habit. Incorporating the cultural insights of local interpreters or patrol reports from junior personnel rarely poses security risks. Yet such perspectives are often systematically undervalued in system design. Which knowledge exclusions are operationally necessary and which reflect unexamined assumptions about whose knowledge counts as legitimate intelligence? 

The cost of exclusion is well documented. In Afghanistan, post-campaign analysis traced intelligence failures to Western-centric assumptions. NATO underestimated the Taliban’s rural mobilisation, partly due to limited engagement with local knowledge which resulted in strategic surprise and reduced adaptability \citep{rietjens2023}. In Nigeria, the military has acknowledged that poor intelligence contributed to civilian deaths in airstrikes \citep{ogbozor2025}. These failures illustrate how narrow epistemic inputs can undermine both mission effectiveness and civilian protection. Equity Bias offers a framework to address such blind spots by broadening epistemic participation in AI development. 

\begin{itemize}
    \item \textbf{Equity Archaeology}: Audit which knowledge is embedded in existing systems. Are threat models based solely on classified sources, or do they also incorporate insights from local communities and frontline units?
    \item \textbf{Co-Creating Meaning}: Redesign AI systems in collaboration with underrepresented knowledge holders. This improves ground-truth validation, supports culturally informed analysis, and deepens understanding of local population dynamics.
    \item \textbf{Ongoing Accountability}: Establish feedback loops that compare AI predictions with real-world outcomes, enabling systems to learn and adapt as conditions evolve.
\end{itemize}

Equity Bias reshapes defence AI decision-making. Incorporating multiple knowledge sources enhances situational awareness, improves threat assessment, and strengthens adaptability. Epistemic inclusion is not just an ethical consideration. It can be a strategic asset that produces more effective defence systems.

\section{Discussion}

Equity Bias challenges conventional views of efficiency in AI development. Traditional metrics emphasise speed, scale, and technical performance, often seeing diverse perspectives as procedural obstacles. This narrow focus hides the full costs of excluding knowledge: missed signals, system failures, and eroding public trust. 

Instead, Equity Bias redefines efficiency as the ability of AI systems to learn from and adapt to a broad range of knowledge. Excluding underrepresented perspectives often causes more problems than including them. For example, climate models that ignore Indigenous knowledge may miss important ecosystem changes. Financial algorithms designed for wealthy markets may misjudge creditworthiness in other economic contexts. Prioritising this broader knowledge adaptability leads to AI systems that are more robust and effective over time. 

\subsection{The Scale-Context Paradox in Practice}

A persistent tension in AI governance is often framed as a trade-off between scale and context: systems designed to operate across populations are assumed to be incompatible with situated, context-specific knowledge. This framing obscures the underlying issue. The core tension is not scale versus context, but uniformity versus adaptability.

Machine learning systems scale by detecting patterns across large datasets, yet many meaningful signals are locally situated and resist broad generalisation. Environmental models trained on temperate climates can miss pollution patterns specific to tropical regions, while educational AI optimised for standardised testing may misinterpret learning in culturally diverse classrooms. These failures arise not from scale itself, but from the assumption that systems must behave uniformly across contexts.

True scalability requires engaging with difference rather than erasing it. Systems that scale effectively do so through structured adaptation, in which shared principles and infrastructures enable contextual variation. Equity Bias operationalises this approach by grounding implementation in local knowledge while maintaining coherence through a common methodological framework. The AI Life Cycle provides a consistent process through which stakeholders determine which system features require standardisation for technical coherence and which must remain adaptable to local conditions.

The dengue surveillance case illustrates this dynamic. Models trained solely on institutional data missed early outbreak signals circulating through social media and community reporting. When local data sources were integrated, the system achieved both wider coverage and greater contextual accuracy. Scalability emerged through adaptive integration rather than uniform data practices.

By designing for adaptability at scale, Equity Bias reframes the scale–context paradox as a productive space for negotiation rather than a problem to be resolved through standardisation. Systems that ignore local knowledge become brittle, while those that abandon scalability forfeit broader impact.

\subsection{Power, Inequality, and Epistemic Governance}

Epistemic exclusion is not merely conceptual but materially grounded in economic inequality. Communities without adequate technical infrastructure, educational opportunities, or institutional access are systematically excluded from shaping AI systems. At the same time, AI development remains concentrated within affluent institutions such as elite universities and technology firms, embedding their social contexts into systems deployed globally.

This material inequality translates directly to epistemic power. AI shapes power by deciding what knowledge counts and whose experiences are recognised. Epistemic power shapes how reality is defined, what problems are considered worthy of attention, and which solutions appear viable. Decisions about data collection, model design, and evaluation are acts of epistemic governance. Often made by experts within closed institutions, these choices concentrate knowledge-making power among those who benefit from existing systems.

Consider some automated resume screening systems. Built on hiring data from elite firms, they encode the epistemic assumption that a 'qualified candidate' means specific educational pedigrees and career trajectories. When deployed widely, these systems can exclude candidates with non-traditional paths such as career changers, self-taught developers, and those with employment gaps. The failure is not technical. The algorithms work as designed. The failure is epistemic. Systems cannot recognise forms of qualification outside their training inputs. Equity Archaeology would reveal whose conception of 'qualified' shaped the system before deployment.

Economically disadvantaged people are often the most exposed to the impact of AI yet are systematically excluded from shaping it. Marginalised groups lack avenues to influence AI and face the devaluation of their knowledge in the systems that increasingly govern their lives. Without addressing these structural barriers, AI risks deepening social divides rather than helping bridge them.

Equity Bias counters these power imbalances by broadening whose knowledge shapes AI. Intentional investments and a redistribution of power are required to increase access to AI development. Much of the technology sector recognises the need for public engagement but does not know where or how to engage. By making visible whose knowledge is included and excluded, Equity Bias helps translate ethical principles into practical AI design. Only through addressing both material conditions and institutional power structures can AI systems reflect the full complexity of human experience and serve all users equitably.

\subsection{Practical Implications}

Implementing Equity Bias requires institutional change supported by material resources and long-term investment. Successfully recognising, translating, and integrating multiple sources of knowledge depends on infrastructure, skilled personnel, and sustained relationships with communities.

Critics may argue that Equity Bias slows development. This framing misunderstands where delays actually occur. Rapid deployment of epistemically brittle systems often leads to costly post-deployment failures. These include redesigns, legal challenges, regulatory interventions, and resistance from communities who reject systems built without them. The OFQUAL grading system example illustrated this point. It was developed quickly, yet its failure required complete abandonment and restart. Upfront epistemic investment prevents such failures. Investing time during design when adaptation is possible, rather than after deployment, helps prevent entrenched systems that may fail in public.

Inclusion is often viewed as a drag on efficiency. In reality, the real inefficiencies lie in exclusions. System failures require costly redesigns. Discriminatory outcomes create legal risks. Public backlash delays deployment. Loss of trust limits adoption. A more complete economic assessment must weigh these hidden costs against the investments needed to build inclusive systems. Inclusive AI can also deliver strategic benefits by improving output accuracy and expanding reach across diverse populations.

Successful implementation of Equity Bias is grounded in robust processes. Equity archaeology should identify whose knowledge shaped the system, ensure affected communities help define problems and evaluation criteria, and enable challenges to underlying assumptions. Success requires transparency about interpretive positions and accountability through contestability rather than claims of neutrality or universal fairness.

Fostering epistemic inclusion does not eliminate leadership or expertise. Clinical, military, and policy decisions still require clear authority. The challenge is to integrate a wider range of insights without falling into indecision. Balancing coherence and openness strengthens decision-making rather than undermining it.

For policymakers, Equity Bias shows how including multiple knowledge sources can improve AI’s accuracy, fairness, and trustworthiness. It guides decisions on funding, regulation, and design by making knowledge inclusion central to building reliable AI that serves all communities. Regulatory frameworks should require transparency about whose knowledge shapes systems and establish mechanisms for contestation. This shifts the focus from impossible neutrality to achievable accountability.

\subsection{Limitations}

Equity Bias recognises its own boundaries. Inclusion and dialogue may be constrained by deeply conflicting values between communities. Practical barriers such as limited time, resources, or institutional support also pose a risk. While the framework advocates for broadening whose knowledge shapes AI, it does not claim that all perspectives are equally relevant in every context. Navigating this balance can be difficult, particularly for institutions used to narrow definitions of expertise. 

A core challenge is discerning which knowledge is most applicable without reverting to exclusionary criteria. Equity Bias aims to expand legitimate epistemic inputs while supporting rigorous, context-sensitive evaluation. The framework's reframing of bias as unavoidable and potentially productive may challenge those committed to neutrality as an achievable goal.

Emphasis on ongoing stakeholder engagement requires sustained investment in time, personnel, and infrastructure. Organisations with limited resources may struggle to implement meaningful participation, potentially creating a two-tier system. Well-funded institutions may be able to pursue epistemic inclusion while others cannot, thereby reproducing the very inequalities the framework seeks to address. This raises questions about equitable access to Equity Bias itself. 

There is a risk that inclusion efforts may be co-opted for appearances rather than substance. Without structural shifts in power and resource distribution, epistemic inclusion may remain symbolic rather than transformative. Equity Bias treats disagreement as productive, yet some conflicts may prove genuinely irreconcilable. When stakeholder groups hold fundamentally incompatible worldviews, continued negotiation may itself become untenable.  

\subsection{Future Research Directions}

Future research should examine how Equity Bias affects AI performance across sectors and settings. Under what conditions does the AI Life Cycle succeed or fail? How do different institutional structures enable or constrain epistemic inclusion? Do participatory processes produce measurably different outcomes compared to conventional approaches? Developing practical tools to assess and build epistemic capacity within organisations is essential. 

Emphasis on stakeholder engagement raises practical questions about implementation. Practical implementation requires tools for conducting Equity Archaeology, facilitating co-design across epistemic differences, and establishing meaningful accountability mechanisms.

Economic analysis is needed to map the full costs and benefits of epistemic inclusion. This includes assessing hidden costs of exclusion (system failures, legal challenges, public backlash) against investments required for participatory design. How are these costs distributed across organisations of different sizes and resource levels, and what support structures could enable broader adoption?

Policy research may explore how governance frameworks could support epistemic inclusion at scale. How can regulation incentivise or mandate broader knowledge participation? What role might international standards, professional accreditation, or public procurement requirements play in embedding these practices? Empirical validation across multiple contexts will determine whether Equity Bias can deliver on its promise of more equitable, effective, and accountable AI. 

\section{Conclusion}

Equity Bias offers a path forward by rejecting the comfortable myth of neutral AI. For policymakers and technology companies alike, this means moving beyond technical fixes towards governance that examines the epistemic foundations of AI systems. Equity Bias does not claim to be the only path forward. It can work alongside other existing critical frameworks and those yet to emerge. Just as it advocates for multiple knowledge systems within AI, Equity Bias positions itself as one voice in a pluralistic conversation about how to build more equitable technologies. 

We have a clear choice. The world can continue to build systems that quietly reproduce existing inequalities under the banner of objectivity. Or it can engage in the more difficult but necessary task of negotiating whose knowledge counts in the technologies shaping our lives. Bias exists. Equity Bias offers a framework for negotiating bias towards AI for the many, not just the few.

\bibliography{references}

@article{bakdash2023,
  author  = {Bakdash, L. and Abid, A. and Gourisankar, A. and Henry, T. L.},
  title   = {Chatting Beyond ChatGPT: Advancing Equity Through AI-Driven Language Interpretation},
  journal = {Journal of General Internal Medicine},
  year    = {2023},
  volume  = {39},
  number  = {3},
  pages   = {492--495},
  doi     = {10.1007/s11606-023-08497-6}
}

@book{benjamin2020,
  author    = {Benjamin, Ruha},
  title     = {Race After Technology: Abolitionist Tools for the New Jim Code},
  publisher = {Polity},
  address   = {Cambridge},
  year      = {2020}
}

@article{birhane2021,
  author  = {Birhane, Abeba},
  title   = {Algorithmic injustice: a relational ethics approach},
  journal = {Patterns},
  year    = {2021},
  volume  = {2},
  number  = {2},
  pages   = {100205},
  doi     = {10.1016/j.patter.2021.100205}
}

@article{buijsman2023,
  author  = {Buijsman, S.},
  title   = {Navigating fairness measures and trade-offs},
  journal = {AI and Ethics},
  year    = {2023},
  volume  = {4},
  number  = {4},
  pages   = {1323--1334},
  doi     = {10.1007/s43681-023-00318-0}
}

@article{catania2023,
  author  = {Catania, B. and Guerrini, G. and Accinelli, C.},
  title   = {Fairness \& friends in the data science era},
  journal = {AI and Society},
  year    = {2023},
  volume  = {38},
  number  = {2},
  pages   = {721--731},
  doi     = {10.1007/s00146-022-01472-5}
}

@article{chen2023,
  author  = {Chen, Z.},
  title   = {Ethics and discrimination in artificial intelligence-enabled recruitment practices},
  journal = {Humanities and Social Sciences Communications},
  year    = {2023},
  volume  = {10},
  number  = {1},
  pages   = {1--12},
  doi     = {10.1057/s41599-023-02079-x}
}

@article{chouldechova2017,
  author  = {Chouldechova, Alexandra},
  title   = {Fair Prediction with Disparate Impact: A Study of Bias in Recidivism Prediction Instruments},
  journal = {Big Data},
  year    = {2017},
  volume  = {5},
  number  = {2},
  pages   = {153--163},
  doi     = {10.1089/big.2016.0047}
}

@article{colongonzalez2023,
  author  = {Col{\'o}n-Gonz{\'a}lez, F. J. and Gibb, R. and Khan, K. and Watts, A. and Lowe, R. and Brady, O.},
  title   = {Projecting the future incidence and burden of dengue in Southeast Asia},
  journal = {Nature Communications},
  year    = {2023},
  volume  = {14},
  number  = {1},
  pages   = {1--10},
  doi     = {10.1038/s41467-023-41017-y}
}

@book{costanzachock2020,
  author    = {Costanza-Chock, Sasha},
  title     = {Design Justice: Community-Led Practices to Build the Worlds We Need},
  publisher = {MIT Press},
  address   = {Cambridge, MA},
  year      = {2020}
}

@article{cote2024,
  author  = {C{\^o}t{\'e}, C. I.},
  title   = {A critical and systematic literature review of epistemic justice applied to healthcare: recommendations for a patient partnership approach},
  journal = {Medicine, Health Care and Philosophy},
  year    = {2024},
  volume  = {27},
  number  = {3},
  pages   = {455--477},
  doi     = {10.1007/s11019-024-10210-1}
}

@article{crenshaw1989,
  author  = {Crenshaw, Kimberl{\'e}},
  title   = {Demarginalizing the Intersection of Race and Sex},
  journal = {University of Chicago Legal Forum},
  year    = {1989},
  volume  = {1989},
  url     = {http://chicagounbound.uchicago.edu/uclf/vol1989/iss1/8},
  note    = {Accessed September 27, 2025}
}

@book{dignazio2020,
  author    = {D'Ignazio, Catherine and Klein, Lauren F.},
  title     = {Data Feminism},
  publisher = {MIT Press},
  address   = {Cambridge, MA},
  year      = {2020}
}

@article{drage2022,
  author  = {Drage, Emma and Mackereth, Kirsten},
  title   = {Does AI Debias Recruitment? Race, Gender, and AI's ``Eradication of Difference''},
  journal = {Philosophy \& Technology},
  year    = {2022},
  volume  = {35},
  pages   = {89},
  doi     = {10.1007/s13347-022-00543-1}
}

@article{ferrara2024,
  author  = {Ferrara, Emilio},
  title   = {The Butterfly Effect in artificial intelligence systems: Implications for AI bias and fairness},
  journal = {Machine Learning with Applications},
  year    = {2024},
  volume  = {15},
  pages   = {2362--2369},
  doi     = {10.1016/j.mlwa.2024.100525}
}

@book{frank2013,
  author    = {Frank, Arthur W.},
  title     = {The Wounded Storyteller},
  edition   = {2},
  publisher = {University of Chicago Press},
  address   = {Chicago},
  year      = {2013}
}

@book{fricker2007,
  author    = {Fricker, Miranda},
  title     = {Epistemic Injustice: Power and the Ethics of Knowing},
  publisher = {Oxford University Press},
  address   = {Oxford},
  year      = {2007}
}

@book{gadamer2004,
  author    = {Gadamer, Hans-Georg},
  title     = {Truth and Method},
  edition   = {2},
  publisher = {Bloomsbury Academic},
  address   = {London},
  year      = {2004}
}

@article{gichoya2023,
  author  = {Gichoya, J. W. and Thomas, K. and Celi, L. and Safdar, N. and Banerjee, I. and Banja, J. and others},
  title   = {AI pitfalls and what not to do: mitigating bias in AI},
  journal = {British Journal of Radiology},
  year    = {2023},
  volume  = {96},
  number  = {1150},
  doi     = {10.1259/bjr.20230023}
}

@article{jui2024,
  author  = {Jui, T. and Rivas, P.},
  title   = {Fairness issues, current approaches, and challenges in machine learning models},
  journal = {International Journal of Machine Learning and Cybernetics},
  year    = {2024},
  volume  = {15},
  number  = {8},
  pages   = {3095--3125},
  doi     = {10.1007/s13042-023-02083-2}
}

@article{jungherr2023,
  author  = {Jungherr, Andreas},
  title   = {Artificial Intelligence and Democracy: A Conceptual Framework},
  journal = {Social Media and Society},
  year    = {2023},
  volume  = {9},
  number  = {3},
  doi     = {10.1177/20563051231186353}
}

@inproceedings{kay2024,
  author    = {Kay, Jack and Kasirzadeh, A. and Mohamed, S.},
  title     = {Epistemic Injustice in Generative AI},
  booktitle = {Proceedings of the AAAI/ACM Conference on AI, Ethics, and Society},
  year      = {2024},
  volume    = {7},
  number    = {1},
  pages     = {684--697},
  doi       = {10.1609/aies.v7I1.31671}
}

@article{kelan2024,
  author  = {Kelan, Elisabeth K.},
  title   = {Algorithmic inclusion: Shaping the predictive algorithms of artificial intelligence in hiring},
  journal = {Human Resource Management Journal},
  year    = {2024},
  volume  = {34},
  number  = {3},
  pages   = {694--707},
  doi     = {10.1111/1748-8583.12511}
}

@article{lindemann2024,
  author  = {Lindemann, N. F.},
  title   = {Chatbots, search engines, and the sealing of knowledges},
  journal = {AI and Society},
  year    = {2024},
  volume  = {40},
  pages   = {5063--5076},
  doi     = {10.1007/s00146-024-01944-w}
}

@inproceedings{lopezperez2025,
	author    = {L{\'o}pez-P{\'e}rez, M. and Hauberg, S. and Feragen, A.},
	title     = {Are Generative Models Fair? A Study of Racial Bias in Dermatological Image Generation},
	booktitle = {Image Analysis. SCIA 2025. Lecture Notes in Computer Science},
	editor    = {Petersen, J. and Dahl, V. A.},
	volume    = {15726},
	pages     = {389--402},
	year      = {2025},
	publisher = {Springer, Cham},
	doi       = {10.1007/978-3-031-95918-9_27},
	url       = {https://link.springer.com/chapter/10.1007/978-3-031-95918-9_27}
}

@incollection{madaio2021,
  author    = {Madaio, M. and Blodgett, S. and Mayfield, E. and Dixon-Rom{\'a}n, E.},
  title     = {Beyond "Fairness": Structural (In)justice Lenses on AI for Education},
  booktitle = {The Ethics of Artificial Intelligence in Education},
  publisher = {Taylor and Francis},
  year      = {2021},
  doi       = {10.4324/9780429329067-11}
}

@book{noble2018,
  author    = {Noble, Safiya Umoja},
  title     = {Algorithms of Oppression},
  publisher = {NYU Press},
  address   = {New York},
  year      = {2018},
  doi       = {10.2307/j.ctt1pwt9w5}
}

@article{ogbozor2025,
  author  = {Ogbozor, E.},
  title   = {Counterinsurgency airstrike mishap, intelligence failure and civilian harm in northern Nigeria},
  journal = {Journal of Policing, Intelligence and Counter Terrorism},
  year    = {2025},
  volume  = {20},
  number  = {2},
  pages   = {193--208},
  doi     = {10.1080/18335330.2025.2449675}
}

@article{pessach2022,
  author  = {Pessach, Dana and Shmueli, Erez},
  title   = {A Review on Fairness in Machine Learning},
  journal = {ACM Computing Surveys},
  year    = {2022},
  volume  = {55},
  number  = {3},
  pages   = {1--44},
  doi     = {10.1145/3494672}
}

@article{ramadona2019,
  author  = {Ramadona, A. L. and Tozan, Y. and Lazuardi, L. and Rockl{\"o}v, J.},
  title   = {A combination of incidence data and mobility proxies from social media predicts the intra-urban spread of dengue},
  journal = {PLOS Neglected Tropical Diseases},
  year    = {2019},
  volume  = {13},
  number  = {4},
  pages   = {e0007298},
  doi     = {10.1371/journal.pntd.0007298}
}

@article{rietjens2023,
  author  = {Rietjens, Sebastiaan},
  title   = {NATO’s Struggle for Intelligence in Afghanistan},
  journal = {Armed Forces and Society},
  year    = {2023},
  volume  = {49},
  number  = {4},
  pages   = {1001--1012},
  doi     = {10.1177/0095327x221116138}
}

@inproceedings{simson2024,
  author    = {Simson, J. and Fabris, A. and Kern, C.},
  title     = {Lazy Data Practices Harm Fairness Research},
  booktitle = {Proceedings of the 2024 ACM Conference on Fairness, Accountability, and Transparency},
  year      = {2024},
  pages     = {642--649},
  doi       = {10.1145/3630106.3658931}
}

@article{song2024,
  author  = {Song, Y. and Weisberg, L. R. and Zhang, S. and Tian, X. and Boyer, K. and Israel, M.},
  title   = {A framework for inclusive AI learning design for diverse learners},
  journal = {Computers and Education: Artificial Intelligence},
  year    = {2024},
  volume  = {6},
  pages   = {100212},
  doi     = {10.1016/j.caeai.2024.100212}
}

@article{straw2022,
  author  = {Straw, I. and Wu, H.},
  title   = {Investigating for bias in healthcare algorithms},
  journal = {BMJ Health \& Care Informatics},
  year    = {2022},
  volume  = {29},
  number  = {1},
  pages   = {100457},
  doi     = {10.1136/bmjhci-2021-100457}
}

@article{veale2017,
  author  = {Veale, Michael and Binns, Reuben},
  title   = {Fairer machine learning in the real world},
  journal = {Big Data and Society},
  year    = {2017},
  volume  = {4},
  number  = {2},
  doi     = {10.1177/2053951717743530}
}

@inproceedings{wang2023,
  author    = {Wang, H. and He, L. and Gao, R. and Calmon, F. P.},
  title     = {Aleatoric and Epistemic Discrimination},
  booktitle = {Advances in Neural Information Processing Systems},
  year      = {2023},
  volume    = {36},
  pages     = {27040--27062},
  doi       = {10.5555/3666122.3667298}
}

@article{yang2024,
  author  = {Yang, Y. and Lin, M. and Zhao, H. and Peng, Y. and Huang, F. and Lu, Z.},
  title   = {A survey of recent methods for addressing AI fairness and bias in biomedicine},
  journal = {Journal of Biomedical Informatics},
  year    = {2024},
  volume  = {154},
  pages   = {104646},
  doi     = {10.1016/j.jbi.2024.104646}
}

@article{zajko2022,
  author  = {Zajko, Mike},
  title   = {Artificial intelligence, algorithms, and social inequality},
  journal = {Sociology Compass},
  year    = {2022},
  volume  = {16},
  number  = {3},
  pages   = {e12962},
  doi     = {10.1111/soc4.12962}
}

\end{document}